\documentclass[12pt,epsfig]{article}
\usepackage{graphicx}
\usepackage{graphics}
\usepackage[numbers]{natbib}

\def\mathnew{\mathsurround=0pt}
\def\simov#1#2{\lower .5pt\vbox{\baselineskip0pt \lineskip-.5pt
       \ialign{$\mathnew#1\hfil##\hfil$\crcr#2\crcr\sim\crcr}}}
\def\simg{\mathrel{\mathpalette\simov >}}
\def\siml{\mathrel{\mathpalette\simov <}}

\def\Mesz{M\'esz\'aros\,}

\def\msun{M_\odot}

\def\grs{$\gamma$-rays~}

\def\vareps{\varepsilon}

\def\noind{\noindent}

\def\beq{\begin{equation}}
\def\enq{\end{equation}}
\def\bea{\begin{eqnarray}}
\def\ena{\end{eqnarray}}
\def\bec{\begin{center}}
\def\enc{\end{center}}

\def\blist{\begin{list}{$\bullet$}{\itemsep 0.0in \parsep 0.0in}}
\def\elist{\end{list}}
\def\bitem{\begin{list}{\arabic{enumi}.}{\usecounter{enumi} \itemsep 0.0in \parsep 0.0in}}
\def\eitem{\end{list}}
\def\cm{\hbox{~cm}}
\def\s{\hbox{~s}}
\def\erg{\hbox{~erg}}

\def\GeV{\hbox{~GeV}}
\def\MeV{\hbox{~MeV}}

\def\eV{\hbox{~eV}}

\def\part{\partial}

\def\barnue{{\bar \nu_e}} 
\def\numu{\nu_\mu}
\def\barnumu{{\bar \nu}_\mu}

\def\kpc{{\rm kpc}}
\def\Mpc{{\rm Mpc}}
\def\Gpc{{\rm Gpc}}
\def\sr{{\rm sr}}

\def\yr{{\rm yr}}

\def\m-pl{m_{Pl}}

\def\h75{h_{75}}
\def\Omh75{\Omega h^2_{75}}
\def\Omh70{\Omega h^2_{70}}

\def\G{{\rm G}}

\def\fun#1#2{\lower3.6pt\vbox{\baselineskip0pt\lineskip.9pt
  \ialign{$\mathsurround=0pt#1\hfil##\hfil$\crcr#2\crcr\sim\crcr}}}
\def\VEV#1{\left\langle #1\right\rangle}

\def\jcap{Jour. Cosmology and Astro-Particle Phys.\,}
\def\mnras{M.N.R.A.S.\,}

\def\apj{Astrophys.J.\,}
\def\apjl{Astrophys.J.Lett.\,}

\def\nat{Nature\,}

\def\prd{Phys.Rev.D\,}
\def\araa{Annu.Rev.Astron.Astrophys.\,}
\def\aap{Astron.Astrophys.\,}

\begin{document}


\bec
{\bf \Large Astrophysical Sources of High Energy Neutrinos \\ in the IceCube Era\footnote{\bf Preprint - Accepted by Annual Review of Nuclear and Particle Science; to appear in vol. 67 (2017)}}
\enc

\bec P. \Mesz$^1$ \enc
{$^1$Department of Astronomy \& Astrophysics, Department of Physics, 
Center for Particle and Gravitational Astrophysics,
Pennsylvania State University, University Park, PA 16802, U.S.A.; email: nnp@psu.edu}

\begin{abstract}
High energy neutrino astrophysics has come of age with the discovery by IceCube of 
neutrinos in the TeV to PeV energy range attributable to extragalactic sources at
cosmological distances. At such energies, astrophysical neutrinos must have their
origin in cosmic ray interactions, providing information about the sources of high energy 
cosmic rays, as well as leading to the co-production of high energy gamma-rays. The
intimate link with these latter two independently observed types of radiation
provides important tools in the search for identifying and understanding the nature of 
their astrophysical sources. These neutrinons can provide important constraints about the 
cosmic ray acceleration process, and since they travel essentially unimpeded they can probe 
our Universe out to the farthest cosmological distances.
\end{abstract}






\section{Very High Energy Neutrino Observations}
\label{sec:intro}

The exciting discovery \cite{IC3+13pevnu1,IC3+13pevnu2,IC3+15tevnu} of a diffuse flux of TeV to $\simg$ PeV 
neutrinos of undoubted astrophysical origin was achieved with the cubic kilometer IceCube Cherenkov 
neutrino detector \cite{Gaisser+14-IC3instr}.  This was the culmination of a series of increasingly 
sophisticated experimental developments and observing campaigns, some previous notable milestones in this 
search having been undertaken through the Dumand concept \cite{Babson+90dumand}, the Baikal experiment 
\cite{Avrorin+16baikal} and the Antares experiment \cite{Margiotta+16-ANTARES}.

\begin{figure}[ht]
\begin{minipage}[t]{0.5\textwidth}
\centerline{\includegraphics[width=3in,height=2.3in]{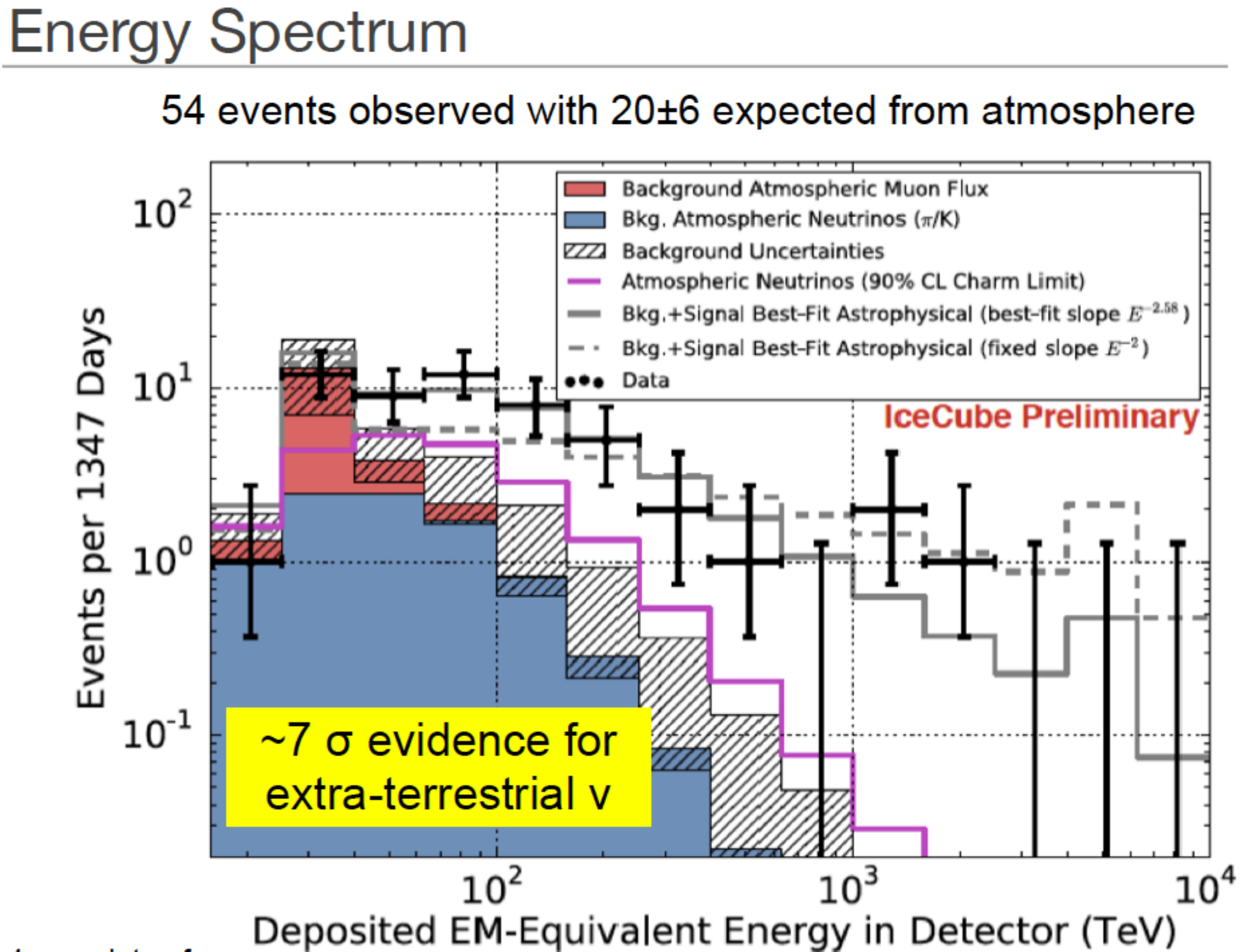}}
\end{minipage}
\vspace*{-1.1in}
\begin{minipage}[t]{0.5\textwidth}
\centerline{\includegraphics[width=3in]{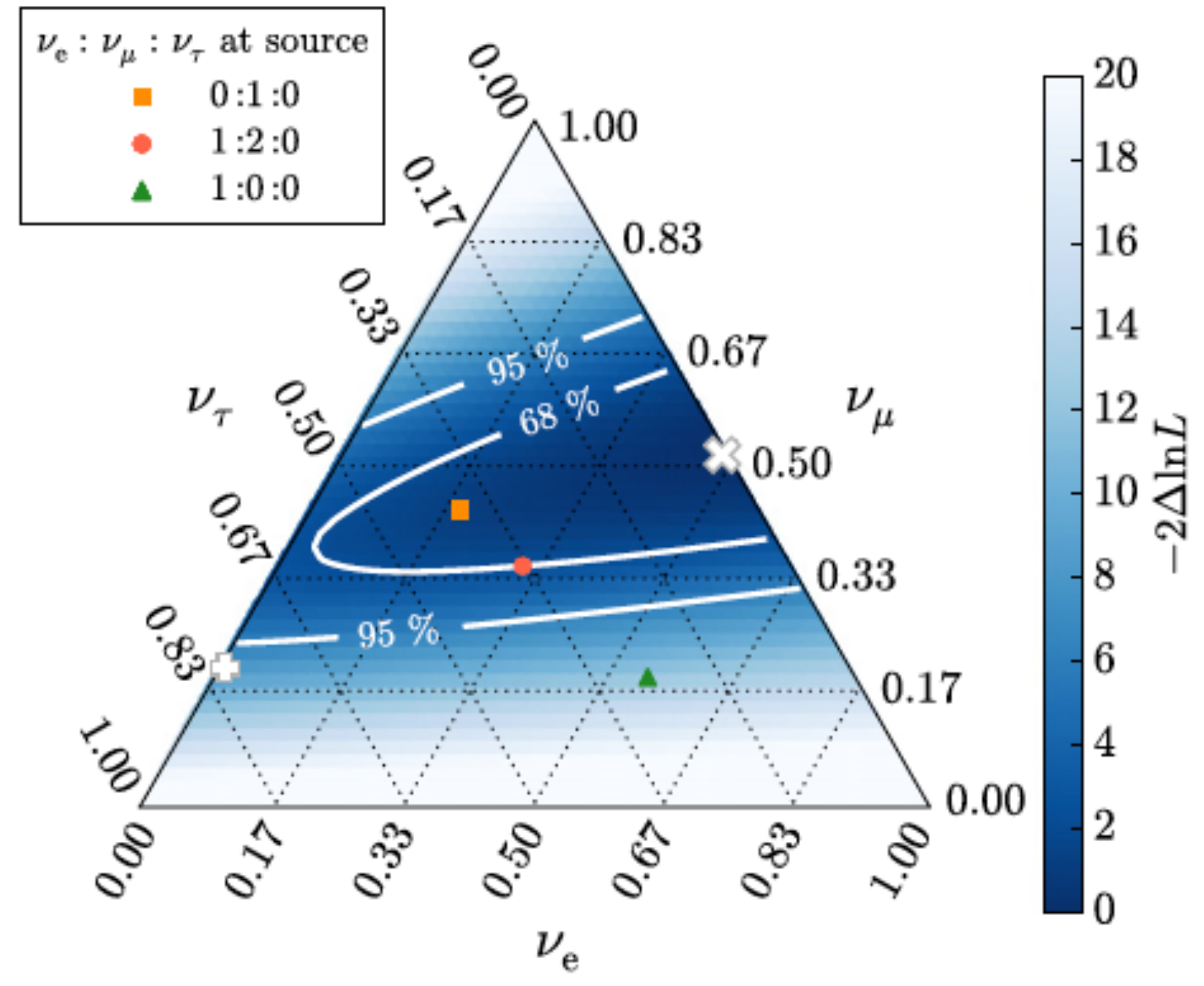}}
\end{minipage}
\vspace*{1.45in}
\caption{\small Left: All-flavor spectrum of VHE neutrinos detected by IceCube 
(from M. Kowalski, in Procs. Neutrino2016, in press; see also \cite{Aartsen+15-IC3nubkggmaxlik}).
Right: Flavor ratio probability distribution of astrophysical neutrinos above 
35 TeV detected by IceCube \cite{Aartsen+15-IC3nubkggmaxlik}.
}
\label{fig:nuspecflav}
\end{figure}

The flux of very high energy (VHE) neutrinos observed by IceCube includes both cascade events ascribed 
to $\nu_e,\barnue$, with angular resolutions $\sim 15-30^o$, and tracks ascribed to $\numu,\barnumu$, 
with angular resolutions $\sim 1^o$. The spectrum clearly departs from that of the atmospheric neutrino 
background with $\geq 7\sigma$ significance (Fig. \ref{fig:nuspecflav}, left).\\

The flavor ratio is compatible with a 1:1:1 distribution as expected from pion decay followed by vacuum 
oscillations across cosmological distances \cite{Aartsen+15-IC3flavor} (Fig. \ref{fig:nuspecflav}, right).
As far as the directions from which the individual neutrinos arrive to Earth, there is no hint of 
a concentration towards either the galactic center or the plane, being compatible with an isotropic
flux \cite{IC3+16-7year}. Neither is there any significant correlation with the arrival direction of 
ultra-high energy cosmic rays detected from the Pierre Auger or the Telescope Array \cite{IC3+16-nucrcorrel}.
While there is no significant correlation with any class of extragalactic objects, the isotropicity 
of the neutrino flux strongly suggests the working assumption that it is of extragalactic origin.

\section{Generic Source Requirements}
\label{sec;generic}

A pre-condition for likely astrophysical VHE neutrino sources is that they also be sources of VHE
cosmic rays, or else that they be irradiated by a flux of cosmic rays from some other source(s).
So far, the isotropicity of the IceCube events has led to the search for possible candidates
concentrating mainly in extragalactic space, e.g. \cite{Murase+16uhenurev}. Such sources could also
be naturally related to the sources of ultra-high energy cosmic rays (UHECRs) observed by the Auger 
and TA cosmic ray arrays \cite{IC3+16-nucrcorrel}, although for the currently detected
maximum neutrino energies of $\siml 3$ PeV it is only necessary to have sources capable of
accelerating CRs up to $\siml 100$ PeV, as discussed below. 

The spectrum shown in Fig. \ref{fig:nuspecflav} can in principle be produced by a CR spectrum
of roughly $N(E_p)\sim E_p^{-2.5}$, steeper then the -2 to -2.2 slope expected from 
first-order Fermi acceleration. The latter, if extending above $\sim 100$ PeV
for the CRs would have resulted in a peak around 6.3 PeV in the electron antineutrino component 
due to the Glashow resonance, e.g. \cite{Murase+13pev, Anchordoqui+13pev, Barger+14glashow, 
Kistler+16multipev, Sahu+16glashow, Anchordoqui+16ic3spbreak, Biehl+16nuglashow}, a resonance 
which is so far not observed, IceCube detecting so far only upper limits above $\sim 3.5$ PeV.
The options are either that the diffuse spectrum is a single power law of slope $\sim -2.5$, or 
else, if the spectrum is flatter below $\sim 6.3$ PeV, there could be a spectral break above a few 
PeV, which could arise naturally in some  systems due to a steepening of the CR diffusion coefficient,
e.g. \cite{Murase+13pev,He+13pevnuhn,Liu+14pevnuhn,Senno+15clugalnu}.

\begin{figure}[h]
\begin{minipage}{0.5\textwidth}
\centerline{\includegraphics[width=3.0in]{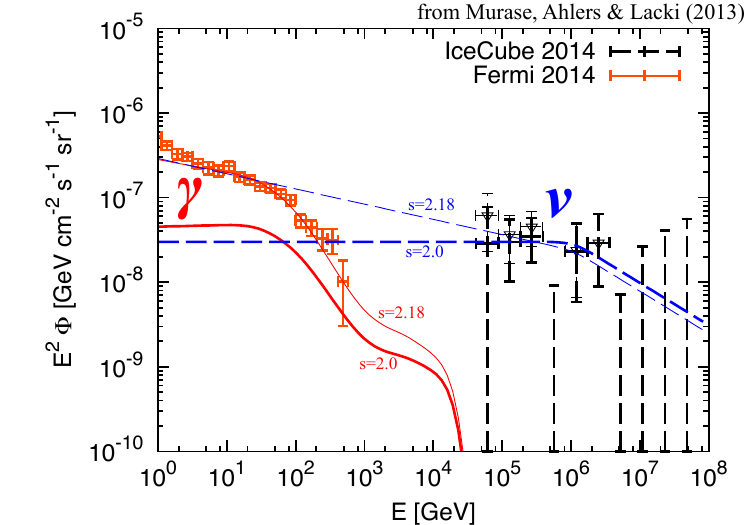}}
\end{minipage}
\begin{minipage}[t]{0.50\textwidth}
\vspace*{-0.9in}
\caption{\small
The isotropic gamma-ray background observed by Fermi compared to the diffuse per-flavor 
neutrino flux observed by IceCube. Black lines: possible neutrino models consistent with 
IceCube data. Red lines: the corresponding $\gamma$-rays of $pp$ scenarios 
reprocessed in the external background light (EBL). The thick and thin solid lines show 
a power-law with slope $s= 2.18$ and $s= 2.0$, resp., with an exponential 
cutoff around PeV.  From \cite{Murase15nuorig}; see also \cite{Ahlers15nulat}.
}
\label{fig:Murase15nuorigfig}
\end{minipage}
\end{figure}

The Fermi diffuse isotropic gamma ray background in the $\sim 10-800\GeV$ photon energy range imposes 
serious constraints on essentially all $pp$ neutrino sources, and on most $p\gamma$ sources as well,
because of the comparable fraction of $\pi^0$ production resulting in secondary TeV to PeV gamma rays 
which cascade against the infrared and microwave extragalactic background light (EBL), ending up down in 
the Fermi range, e.g. Fig. \ref{fig:Murase15nuorigfig}.

The secondary $\gamma$-rays have at birth the same slope as the neutrinos, and the branching ratio of 
charged to neutral pions in $pp$ (or $p\gamma$) interactions implies that the two flux levels are
related through $E_\gamma^2 \Phi_{E_\gamma} =2 E_\nu^2 \Phi_{E_\nu}$. These initial $\gamma$-rays of 
energy $E_\gamma \simg m_e^2/E_{EBL}$ pair-produce against lower energy (infrared) photons from the 
diffuse EBL due to stars, as well as the cosmic microwave background. The pairs then inverse-Compton
scatter against the same low energy photons resulting in new, lower energy $\gamma$-rays, etc.
The resulting electromagnetic (EM) cascades lead to a universal spectral shape \cite{Berezinsky+75nucasc}
which converts all the high energy $\gamma$-rays into lower energy, sub-TeV photons. 
{\it Fermi} is sensitive in the $\sim 0.1-800\GeV$ range \cite{Ackermann+15igbfermi}, and this provides 
strong constraints on models, and in particular on a hadronuclear origin of the neutrinos from any kind 
of sources, e.g.  \cite{Murase+13pev,Liu+13pevnuhn,Liu+14pevnuhn,Chang+14pevnugam,Senno+15clugalnu,
Ando+15nutomopev,Zandanel+15nugamclu,Murase15nuorig,Ahlers15nulat}, see Fig. \ref{fig:Murase15nuorigfig}.

A dominant fraction of the Fermi diffuse extragalactic gamma background is in fact well accounted 
for by unresolved distant  blazars \cite{Ackermann+15igbfermi}, whose gamma-ray emission is most likely 
of leptonic origin, i.e. mechanisms which are not associated with neutrinos. On the other hand, many
of the most commonly considered sources such as active galactic nuclei (AGNs), standard gamma-ray bursts
(GRBs), etc.  would be optically thin, i.e. allow free escape to both neutrinos and $\gamma$-rays, the latter 
being restricted by the Fermi observations. This restriction would not apply, however, to  EM-``hidden" 
sources, e.g. \cite{Murase+16hidden}, in which the $p\gamma$ or $pp$ gamma rays are either absorbed or 
degraded to substantially lower energies, such sources being less (or not at all) constrained by Fermi. 

For sources dominated by $pp$ the efficiency (or optical depth) $\tau_{\gamma\gamma}$ of  $\gamma\gamma$ 
absorption is generally uncorrelated to its $pp$ pion formation efficiency $f_{pp}$, since the first 
depends on the photon target column density and the second on the proton target column density. 
For $p\gamma$ sources, however, there is a generic correlation \cite{Waxman+97grbnu} between 
$\tau_{\gamma\gamma}$ and $f_{p\gamma}$, since both depend on the photon target column density.  
The $p\gamma$ opacity is $\kappa_p \sigma_{p\gamma} \sim 0.7\times 10^{-28}\cm^2$, where $\kappa_p$ is 
the inelasticity, while the $\gamma\gamma$ opacity is $\sigma_{\gamma\gamma}\sim 0.1\sigma_T$, where
$\sigma_T$ is the Thomson cross section, so, e.g. \cite{Murase+16hidden},
$\tau_{\gamma\gamma}\sim \sigma_{\gamma\gamma}/(\kappa_P \sigma_{p\gamma})f_{p\gamma}\sim 10^3 f_{p\gamma}$.
Thus, for moderately efficient $p\gamma$ sources, the high energy $\gamma$-rays will be efficiently degraded. 
The final energy where they reappear after the cascades depends on the target photon energy spectrum. 
In addition, since the $\gamma\gamma$ cross section is close to that of Compton scattering, 
the photons can be trapped and partially thermalize as they diffuse out.  This is important for sources 
such as AGNs, standard GRBs or others which are detected also in the optical, X-ray or MeV range, 
where additional constraints are imposed by stacking analyses of source locations against the error boxes 
of individual detected neutrinos.
Thus, in general, compact high energy sources such as GRBs, especially choked GRBs \cite{Murase+13choked,
Senno+16hidden}, white dwarf mergers \cite{Xiao+16wdmnu}, tidal disruption events \cite{Wang+16tdecrnu,
Senno+16tde}, core AGN sources \cite{Kimura+15nullagn}, etc., by virtue of  their high photon density 
are likelier to suffer photon trapping and efficient $\gamma\gamma$ degradation, making them EM-dim or 
effectively EM-hidden.  As discussed in \S \ref{sec:agn}, another way in which $\nu,\gamma$ sources 
can be EM-hidden in the Fermi range is if they are at high redshifts $z\simg 3-4$ \cite{Chang+16grbnugam, 
Xiao+16nuhn}, the longer pathlength ensuring more efficient absorption in the EBL.

\section{Specific Astrophysical Sources}
\label{sec:sources}

\subsection{Active galactic nuclei (AGNs)}
\label{sec:agn}

The term AGN is applied to a small fraction of all galaxies in which either the galactic nucleus is 
a prominent source of radio or X-ray photons, or else (or in addition) it has bright jets emanating
from the nucleus, which are detectable in radio, optical, X-rays or \grs. These emissions are powered
by accretion onto a central massive black hole (MBH) in the nucleus. 
Average galaxies such as the Milky Way generally also have MBHs at the center, and in about 
$\sim 30\%$ of these the gas accreting into the MBH leads to detectable radio or X-ray emission,
in the range of $10^{43}\erg\s^{-1}$, these being called {\it low luminosity AGNs} (LLAGN). 
AGNs can have luminosities up to 4-5 orders of magnitude higher then LLAGNs, and are sub-divided 
into radio-quiet or RQ AGNs (a misleading term, since they have a dominant radio or X-ray nuclear 
emission, present also in LLAGNs); and radio loud or RL AGNs, which show luminous jets, detected 
mostly in radio but in some cases also in optical, X- and/or gamma-rays. 
The RQ and the RL AGNs represent roughly $10^{-1}$ and $10^{-2}$ of the total galaxy population.
The RL (jet) AGNs are further sub-divided into the so-called FR-I and FR-II types.
The FR-I have irregularly shaped, lower luminosity outer jets extending not far beyond the host 
galaxy, the nearest ones sometimes showing a bright, much straighter inner jet inside the galaxy 
image\footnote{An example being the radio, optical and X-ray jets of the famous nearby M87 galaxy.}.
The FR-II have very extended, high luminosity narrow jets, whose dimension (up to few hundred Kpc) 
can far exceed that of the optical host galaxy. 

{\it Blazars} are a rare sub-class of radio-loud AGNs ($\sim 10^{-5}$ of all galaxies), whose jets 
point close to the line of sight. The jets are relativistic (bulk Lorentz factors $\Gamma_j \sim 
5-30$), implying a large Doppler boost of the jet luminosity, which dominates that of the host 
galaxy and the nucleus.  
Blazars include aligned FR-I AGNs called {\it BL Lac} objects, whose luminosities are not overly 
large, $\sim 10^{44}-10^{45}\erg\s^{-1}$; 
and aligned FR-II AGNs called {\it flat spectrum radio quasars} (FSRQs)\footnote{Quasars, or QSRs, 
are AGNs powered by the most massive MBHs, $10^8-10^{10} \msun$; they are called QSOs when detected only 
in optical, and QSRs when detected in radio.} whose luminosity is much larger, 
$\siml 10^{46}-10^{47}\erg\s^{-1}$ (Fig. \ref{fig:agntype}).
\setcounter{figure}{3}
\begin{figure}[h]
\begin{minipage}{0.5\textwidth}
\centerline{\includegraphics[width=1.0\textwidth,height=2.1in]{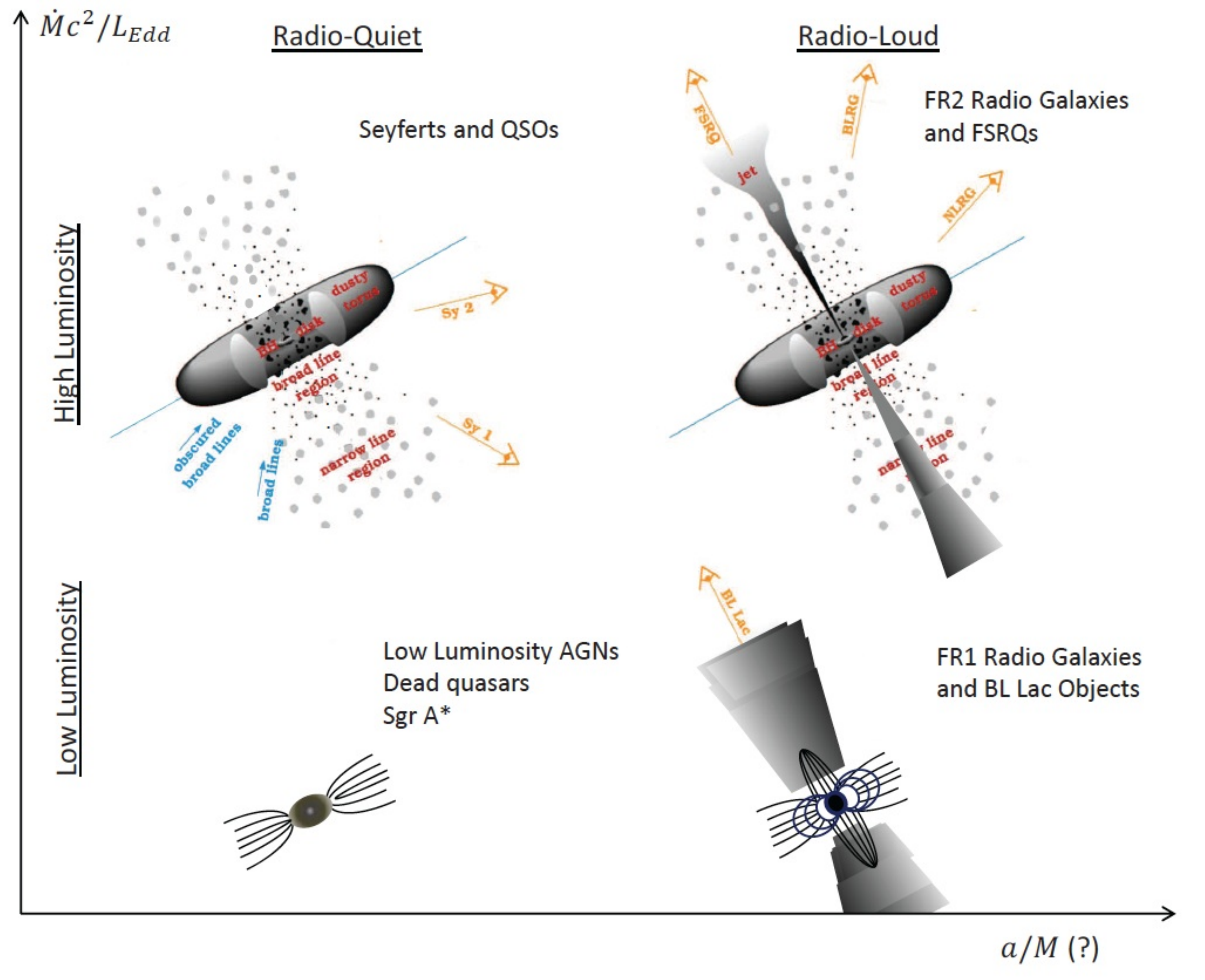}}
\end{minipage}
\begin{minipage}[t]{0.5\textwidth}
\vspace*{-0.7in}
{\small
Sketch of AGN types with relative radio-loudness in the X-axis 
and luminosity in the Y-axis, with an arbitrary division at $10^{45}\erg\s^{-1}$ between 
LL AGNs and high luminosity (HL) AGNs, e.g.  \cite{Dermer+16gamagnrev}. Clockwise from
lower left (LL, RQ) one has LL AGNs, dead quasars, and our own galactic center radio source Sgr A*;
top left (HL, RQ) are Seyfert galaxies and QSOs; top right (RL HL) are FSRQs and FR2 Radio Galaxies;
and bottom right (RL, LL) are Blazars and FR1 Radio Galaxies. 
From \cite{Dermer+16gamagnrev}.  }
\end{minipage}
\label{fig:agntype}
\end{figure}

Models can generically be classified as leptonic, hadronic or lepto-hadronic type, depending on how 
important is the electromagnetic emission of the hadronic secondaries for the observed photon spectra.
In leptonic models it is the primary accelerated electrons which are responsible for the photon spectra, 
even if hadrons were accelerated (their secondary EM emission is negligible). In hadronic models the
hadronic secondary EM emission (or in some cases proton synchrotron) provides the bulk of the 
observed EM spectra, while in lepto-hadronic models it is a mix of both. Protons may be accelerated
in all three types, and the CR efficiency is generally parameterized with the ratio of the luminosities 
in CRs and photons\footnote{Some authors use a ratio of CR luminosity to jet kinetic luminosity 
$\xi'_{CR}=L_{CR}/L_{kin}$.}, $\xi_{CR}=L_{CR}/L_{rad}$

RL AGNs, especially blazars, were the earliest suspected cosmic ray accelerator candidates. Their 
relativistic jets undergo strong shocks as they plow into the intergalactic medium, made detectable 
by the intense non-thermal radiation ascribed to synchrotron and inverse Compton from relativistic
electrons, which are presumed to be accelerated into a power law distribution by a Fermi process 
in the high Mach number shocks. These termination shocks, as well as internal shocks closer in
within  the jet, are also ideal sites for accelerating protons, and could be the source of the 
observed ultra-high high energy CRs (UHECRs) as well as high energy \grs, e.g. 
\cite{Protheroe+83agncrgam,Biermann+87agncr}. 
An additional consequence of this would be VHE neutrinos, e.g. \cite{Stecker+s96nuqsr,Atoyan+01nuagn}. 
This requires that the jet environment provide an adequate column density (or ``optical depth") of 
target photons and/or nucleons. Such targets are undoubtedly present, but the jets combine low target 
densities $n_{t}$ competing with their large dimensions $R$ in producing the optical depth $\tau_{p t} 
\sim n_{t} \sigma_{p t} R$ against proton or photon targets which make pions leading to 
$\nu,\gamma$s, e.g. \cite{Murase+14nuagnrev,Padovani+16extblaznu}.  

In blazars there are four main sources of photons that act as targets for $p\gamma$ interactions, namely:
1) Continuum photons from the optically thick disk accretion disk which feeds the MBH, typically ranging 
over $\sim [ 10-10^5 ]$ eV;
2) Continuum infrared photons from a dusty torus, which often is detected outside the accretion disk, 
typically peaking around  $\sim 10^{-2}-10^{-1}$ eV;
3) Line (H$\alpha \sim 10$ eV) photons from the so-called broad line region (BLR) gas clouds detected 
outside the jet, especially in FSRQs; the BLR also scatters disk and torus continuum photons towards the jet,
4) Nonthermal emission from the inner jet, which in the so-called high-synchrotron peaked (HBL) BL-Lacs, 
which can be detected up to TeV photon energies, is a two-humped spectrum ascribed to synchrotron and 
synchrotron self-Compton (SSC); while in the low-synchrotron peaked (LBL) BL Lacs and FSRQs the two-humps
are best fitted with jet synchrotron accounting for the low energy peak and external Compton accounting
for the higher peak, i.e. scattering by jet electrons of ``external" photons coming from either the 
accretion disk, the dust torus or, in FSRQs,  the BLR's own H$\alpha$ line photons plus the continuum disk
and torus photons which it scatters into the jet.

Besides blazars, other types of AGNs may also contribute to the neutrino background, including
radio-quiet quasars, e.g. \cite{Stecker+s96nuqsr,Alvarez+04agn,Peer+09agn} and LLAGNs 
\cite{Kimura+15nullagn}. The CR acceleration and neutrino emission of such RQ and LLAGN models
is concentrated in the nucleus, where densities are larger. Since $\sigma_{pp}\sim 3\times 10^{-26}\cm^2$ 
while $\sigma_{p\gamma}\sim 10^{28} \cm^2$ near threshold, while the relative increase of the nucleon 
density is larger than that of the photons, the $pp$ interactions can become more important or even
dominant.

\begin{figure}[h]
\begin{minipage}{0.5\textwidth}
\centerline{\includegraphics[width=2.8in]{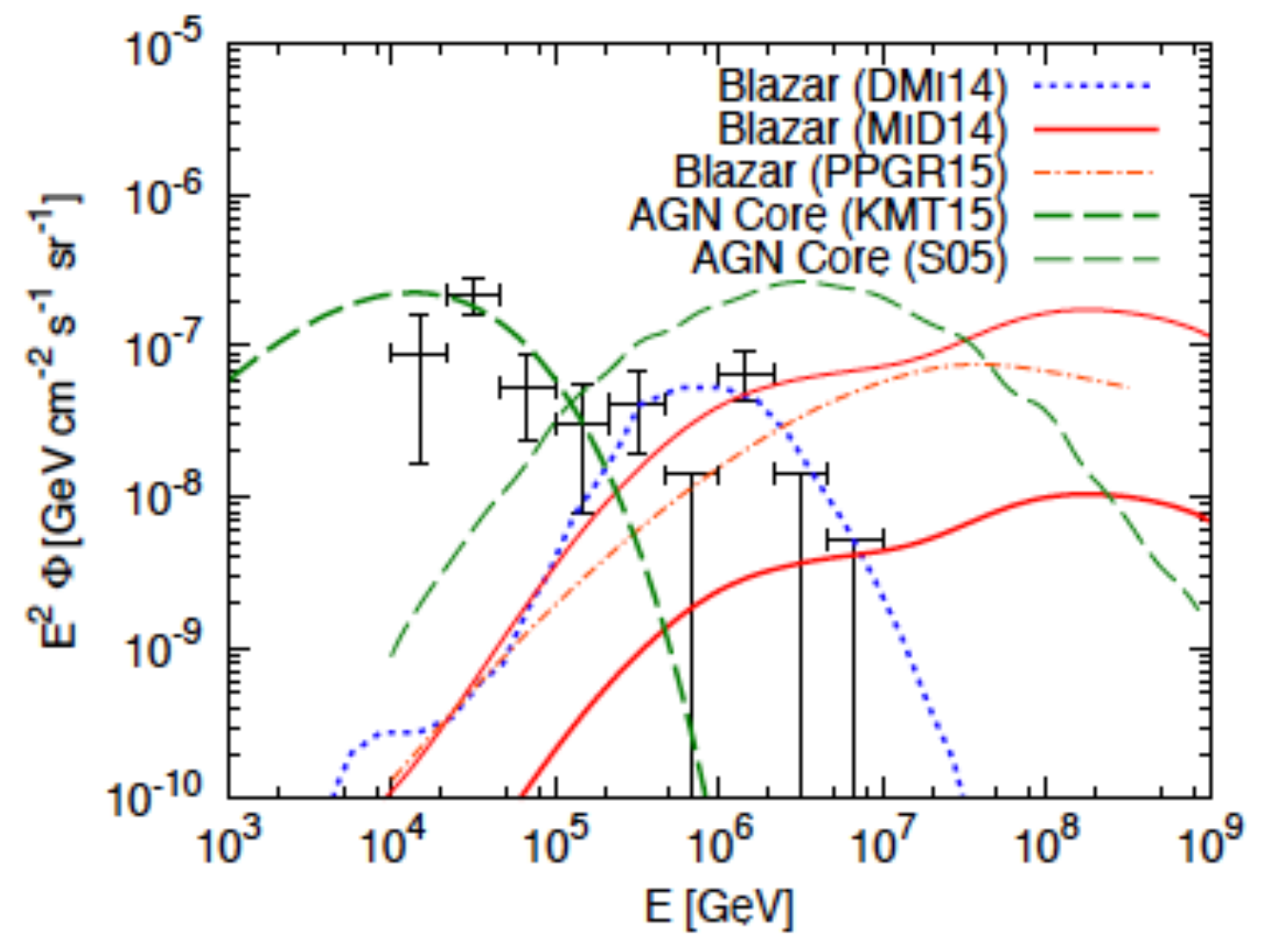}}
\end{minipage}
\begin{minipage}[t]{0.50\textwidth}
\vspace*{-1.0in}
\caption{\small
All-flavor diffuse neutrino intensity for various AGN jet and core models \cite{Murase15nuagnrev}:
DMI14, a FSRQ jet model normalized to the IceCube data at PeV energies assuming $\VEV{z}=2$ 
\cite{Dermer+14fsrqnu}; 
MID14, two different leptonic blazar jet models with high/low CR efficiency \cite{Murase+14nuagnrev}; 
PPGR15, a BL Lac jet model based on a lepto-hadronic scenario \cite{Padovani+15blaznu}; 
KMT15, a LLAGN core model \cite{Kimura+15nullagn}; 
S05, a radio-quiet AGN core model \cite{Stecker13nucoreagn}. 
The diffuse neutrino intensity data is from the IceCube combined likelihood analysis 
\cite{Aartsen+15-IC3nubkggmaxlik}. 
}
\label{fig:nuagnmod}
\end{minipage}
\end{figure}
Fig. \ref{fig:nuagnmod} shows \cite{Murase15nuagnrev} a comparison of the diffuse neutrino 
background predictions from some of the AGN models against IceCube data \cite{Aartsen+15-IC3nubkggmaxlik}.
Typical blazar models have hard spectra, because the $p\gamma$ efficiency increases linearly
with energy, and for a proton slope like -2 or -2.5 the neutrino spectrum has a positive slope. These
models can explain the PeV data but under-predict the TeV data. Also,  these models are in tension with 
the non-observation of a Glashow resonance at 6.3 GeV. The exception is a FSRQ model 
\cite{Dermer+14fsrqnu} where protons are accelerated by a Fermi 2nd order mechanism and the maximum 
proton energy at which acceleration balances escape is $\sim$ 10-100 PeV, while the main targets 
are BLR line photons of $\sim 10$ eV, giving a neutrino cutoff in the few PeV range. 
Older AGN core models have similar problems with the Glashow resonance and also
under-predict the TeV data. A more recent LLAGN core model \cite{Kimura+15nullagn}, assuming a 
radiatively inefficient accretion flow (RIAF) in which $pp$ interactions dominate, reproduces well the 
10-100 TeV data, but under-predicts the PeV data. These are typically one-zone models, which involve
large astrophysical uncertainties, so although they all appear to have difficulties in fitting the 
spectral data they may not necessarily be ruled out. Other, weaker AGNs that have been considered
are radio galaxies, e.g. \cite{Becker+14agnpevnu,Hooper16nuradiogal}. 

An important observational constraint is provided by a recent IceCube study \cite{IC3+16fermiblaznu} 
based on stacking analyses of spatial correlations, which sets limits on the possible cumulative 
contribution of Fermi-2LAC blazars to the diffuse TeV-PeV neutrino flux. They concluded that, assuming 
a -2.5 power law index, they can contribute at most 27\%, or for a -2 index at most 50\% of the total 
observed 10 TeV to 2 PeV neutrino flux, assuming complete oscillation between flavors. Similar results
are obtained by \cite{Neronov+16blaznu}\footnote{Noting that \cite{Resconi+16blaznucr} argue that at a 
$3.3\sigma$ level a bright (HBL) sub-class of blazars could be responsible for some of the IceCube 
neutrinos as well as UHECRs.}.

\subsection{Galaxy clusters/groups and associated sources and shocks}
\label{sec:galclu}


The importance of clusters of galaxies as amplifiers of the  secondary radiation (neutrinos and gamma-rays) 
from intra-cluster UHECR sources was emphasized by \cite{Volk+96clucrgam,Berezinsky+97clucr} and others. 
This is because the CRs, after having been accelerated and undergone some secondary-producing $p\gamma$ 
or $pp$ interactions inside their immediate source of origin (AGNs, supernovae, etc.), they escape into the
intra-cluster medium, which for large clusters of radius $R_{cl}\sim$ few Mpc has typically an average gas 
density $n_0\sim 10^{-3}\cm^{-3}$, magnetic field strength $B_0\sim 10^{-6}\G$ and coherence length 
$\ell_{coh}\sim 30\kpc$. The typical diffusion coefficient $D(E_p)\propto E_P^\alpha$, where $\alpha=
(5/3,1/2)$ for Komolgoroff or Kraichnan turbulence spectra, and the diffusion time out of the cluster 
$t_{esc}\sim R_{cl}^2/6D$ exceeds by orders of magnitude the light crossing time $R_{cl}/c\sim$ 10 Myr. 
During this CR diffusion time their secondary-producing interactions exceed by far those undergone in 
their original source. The interactions after the CRs escape into the intergalactic medium (IGM) are 
typically less important than those undergone within a large cluster.

Accretion of external intergalactic gas onto the cluster gives rise to a stand-off shock, resulting in 
a shocked cluster gas layer and a stationary shock front facing the IGM. Such shocks can accelerate 
electrons to Lorentz factors $\gamma_e\sim 10^7$ which, as they scatter off microwave background photons, 
can contribute \cite{Loeb+00igshock,Keshet+03igshock} to the diffuse extragalactic gamma-ray background.
Cluster accretion shocks are also expected to accelerate CR protons, e.g. \cite{Norman+95clucr,
Inoue+05igshock}, which undergoing photohadronic or hadronuclear interactions also contribute to 
the gamma-ray background, as well as to a diffuse neutrino background \cite{Murase08cluster,
Murase+13pev,Zandanel+15nugamclu}. However, such cluster  accretion shocks are in tension with 
clustering limits \cite{Murase+16uhenurev} and with radio limits \cite{Zandanel+14cluradio}.

Galaxy-galaxy collisions are also expected to occur in clusters of all sizes, all galaxies being thought 
to have undergone at least one (and for large galaxies many) major mergers in their history, in typical
hierarchical growth structure formation schemes, e.g. \cite{Nelson+15pubdataillustris}. 
Single galaxies move in the cluster with virial velocities, and shock-heat the intra-cluster gas,
and also the gas in the colliding galaxy pairs undergoes strong shocks. The kinetic energy input rate  
is comparable to that of the accretion shock onto the cluster \cite{Kashiyama+14pevmerg}.

Fermi acceleration in these various types of shocks can lead to a power law energy distribution of CRs
which are trapped in the cluster for a diffusion time, the latter depending on the shocked layer width, 
magnetic field strength and type of turbulence. For any such sources, the clusters act as CR reservoirs 
\cite{Berezinsky+97clucr, Murase+08clusternu,Kotera+09clucrcomp}, providing for a longer time 
during which they produce secondaries, mainly via $pp$ interactions.  This leads to a neutrino spectrum 
whose slope mimics that of the protons and, assuming a slope $s\sim 2-2.2$, whose diffuse energy 
flux per energy decade $E_\nu^2 \Phi_{E_\nu}$, e.g. \cite{Murase+13pev,Kashiyama+14pevmerg} is 
comparable to that of the first IceCube flux data in the sub-PeV to PeV range \cite{IC3+13pevnu2}. 
One possibility which is allowed by the above mentioned clustering and radio limits is if CR
acceleration occurs in AGNs in clusters and smaller groups of galaxies which serve as CR reservoirs
\cite{Murase+16uhenurev}, the effective density being larger than for accretion shocks since the 
low mass clusters can make a larger contribution.
However, in all cases of optically thin sources (such as the above) proton slopes steeper than $\sim 
- 2.1$ would result in violating the Fermi limits (see, e.g. Fig. \ref{fig:Murase15nuorigfig}).
Here the true diffuse isotropic gamma-ray background is to be understood as the fraction remaining 
after subtraction of the resolved individual sources and the extrapolated contribution of unresolved 
sources \cite{Ackermann+16igbfermi}.

The contributions to the diffuse secondary $\nu,\gamma$ backgrounds from all models will be lower 
\cite{Kotera+09clucrcomp} if the accelerated UHECRs are predominantly heavy elements, as suggested by 
the {\it Auger} observations \cite{Watson16crcomp,Auger+16comp} at energies above $\sim 10^{18}\eV$. 
This is because the individual protons undergoing $p\gamma$ interactions carry only a fraction $1/Z$ 
of the total CR energy, while heavier nuclei are subject to photodesintegration, e.g. 
\cite{Allard+06cosmonu}.

\subsection{Starburst galaxies, supernovae and hypernovae}
\label{sec:sbgsn}

Starburst galaxies (SBGs) are normal galaxies which are undergoing episodes of intense star formation, 
${\dot M}_\ast \sim 1-10~ \msun\yr^{-1}$, lasting $10^6-10^7 \yr$, longer than the lifetime of young 
massive stars, which then become supernovae (SNe). Normal galaxies typically undergo a number of star-forming
episodes since birth, and the steady-state density of galaxies which at any time are starbursts
is $n_{SBG}\sim 3\times 10^{-5}\Mpc^{-3}$, roughly two orders of magnitude less than quiescent galaxies.
About 20-30\% of all star formation in the Universe occurred in such SBGs. 

\begin{figure}[h]
\begin{minipage}{0.5\textwidth}
\centerline{\includegraphics[width=2.8in]{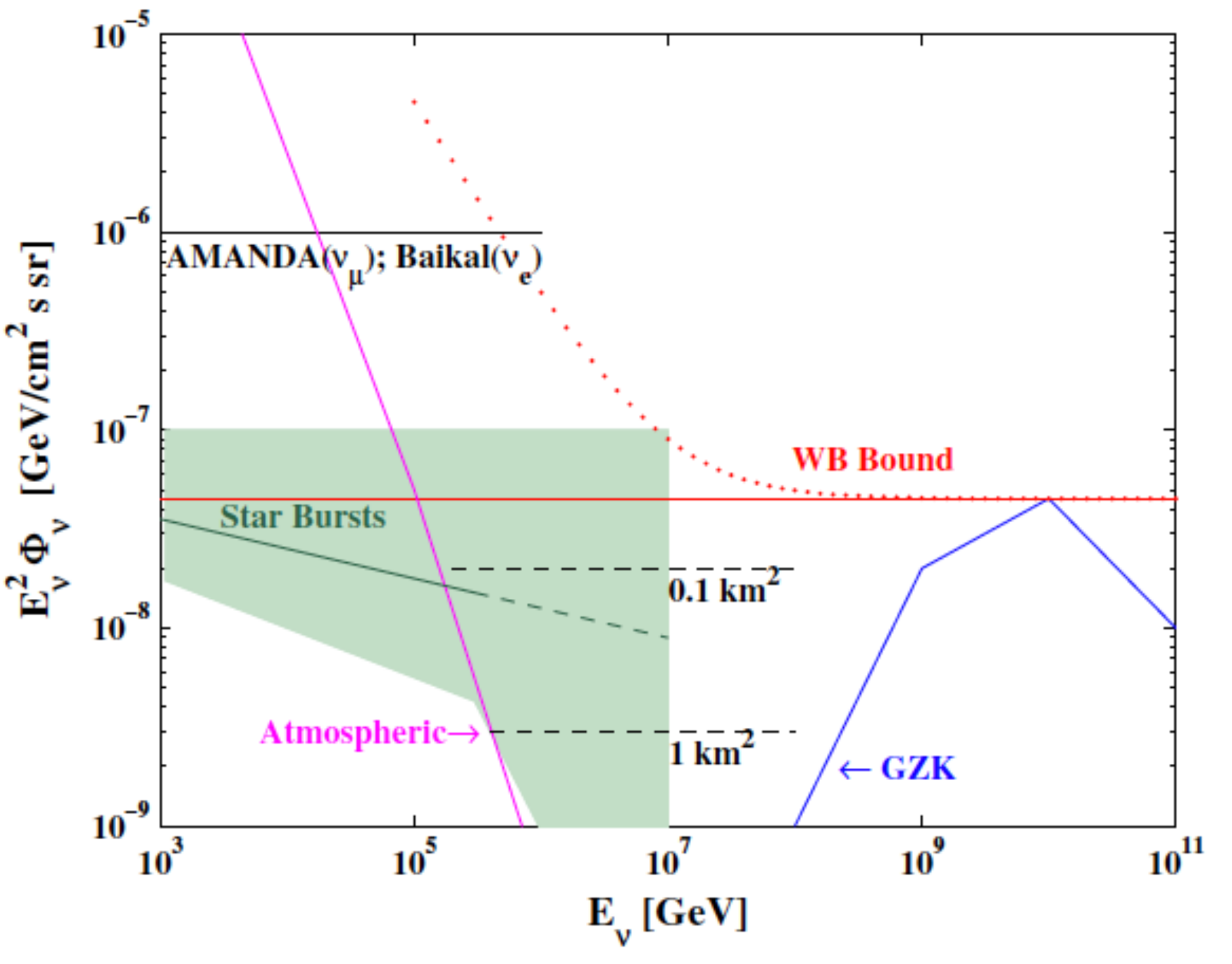}}
\end{minipage}
\begin{minipage}[t]{0.50\textwidth}
\vspace*{-1.0in}
\caption{\small
The possible starburst neutrino background (shaded). The upper boundary is 
for a CR index $s=2$, lower boundary is for $s= 2.25$ for $E_\nu < 10^{14.5}$ eV. 
The solid green line is for $s= 2.15$. Also shown: the WB upper bound;
the neutrino intensity expected from interaction with CMB photons (GZK);
the atmospheric neutrino background; some experimental upper bounds $@$ 2006,
and the approximate sensitivity of 0.1 km$^2$ and 1 km$^2$ optical 
Cherenkov detectors. From \cite{Loeb+06nusbg}.
}
\label{fig:Loeb+06nusbgfg}
\end{minipage}
\end{figure}

Large numbers of SBGs with known redshift distances have measured radio luminosities at 1.4 Ghz,
which is due to synchrotron radiation by relativistic electrons whose cooling time is shorter than 
the SBG phase lifetime. Thus, the energy production rate of electrons is a measure of the radio 
luminosity $L_\omega$ per unit frequency $\omega$, $E_e^2 dN_e/dE_e \simeq 2\omega L_\omega$, where the 
factor $2$ arises because the synchrotron frequency $\omega \propto E_e^2$. In quiescent galaxies like ours
the ratio of energy input in CR protons to electrons is $\eta_{p/e}\sim 50$, but as pointed out by 
\cite{Loeb+06nusbg}, in SBGs the increased SN activity and a  magnetic field  $\simg 10^2$ times larger 
than ours is likely to result in a much slower diffusive escape of the CR protons, which can lose most
of their energy in $pp$ interactions, leading to pions. 
The luminosity per decade of $\nu_\mu +{\bar \nu}_\mu$ is related to the photon luminosity 
$\omega L_\omega$ by $E_\nu dL_\nu =(1/3) \eta_{p/e} E_e^2 d {\dot N}/dE_e = 
(2/3)\eta_{p/e} \omega L_\omega$, where the factor 1/3 is because 2/3 of the proton energy is carried by
charged pions (1/3 by neutral pions), and since the charged pions decay into four particles 
($\pi^+\to \mu^+, {\bar \nu}_\mu \to e^+, \nu_e , \nu_\mu, {\bar \nu}_\mu $), about $1/2$ of the 
charged pions's energy is carried by muon neutrinos. Also, since the secondary electrons 
carry $\sim (2/3)\times 1/4 \sim 1/6$ of the proton energy, in SBGs one expects a CR proton
to electron ratio $\eta_{p/e}\sim 1/6$, so the muon neutrino luminosity per decade of energy is
related to the photon luminosity per decade of frequency by $E_\nu dL/dE_\nu \simeq 4 \omega L_\omega$,
or $E_\nu^2 (d{\dot N}/dE_\nu) \simeq 4 E_\gamma^2 (d{\dot N}/dE_\gamma$).
Thus, on can calibrate the CR luminosities of SBGs, or their expected neutrino energy flux, via their observed 
1.4 GHz radio luminosities. A similar calibration can be also based on an established correlation of the
infrared luminosity to star-formation rate in SBGs.
By analogy with our galaxy's observed CR spectrum below the knee of $N_{obs}(E_p)\propto E^{-2.75}$ 
and its confinement time $t_{esc}\propto E^{-0.6}$, \cite{Loeb+06nusbg} assume a similar inferred 
injection spectrum $N(E_p)\propto E^{-2.15}$ for the SBGs. This predicts a well-motivated
SBG neutrino diffuse flux, shown in fig. \ref{fig:Loeb+06nusbgfg}, which is comparable to the 
Waxman-Bahcall (WB) flux \cite{Waxman+99bound}. A flux of that order is indeed being observed by IceCube 
\cite{IC3+13pevnu2,IC3+15tevnu}, although there is so far no direct evidence linking it to SBGs.

Supernovae are the most likely ultimate sources responsible for accelerating the CRs in SBGs, which
make many more SNe during their starburst phase than normal galaxies. A sub-class of supernovae, called 
hypernovae (HNe), representing a fraction $\sim 5\%$ of the total, are known to occur in all galaxies.
Their ejecta velocities can reach semi-relativistic values, as opposed to $v_{ej}\sim 10^9\cm\s^{-1}$ for 
normal SNe, and their ejecta kinetic energies (isotropic-equivalent value) can reach $\sim 10^{52}\erg$,
as opposed to $\sim 10^{50}-10^{51}\erg$ for SNe. The maximum CR energy achievable by Fermi shock acceleration, 
from eq. (\ref{eq:emaxdsa}), is $E_{max}\simg 10^{15}ZeV$ for SNe and $E_{max} \sim 10^{17}Z\eV$ for HNe.
This was used \cite{Wang+07crhn,Budnik+08hn} for making source-specific neutrino background predictions
before IceCube observations were available, and subsequently, in the light of both IceCube and Fermi data, 
hypernova neutrino production were discussed e.g. by \cite{He+13pevnuhn,Liu+14pevnuhn,
Tamborra+14sbgnugam,Chang+14pevnugam,Bartos+15pevnu}.
A more detailed discussion of the neutrino production and the constraints imposed by Fermi was given 
by \cite{Senno+15clugalnu} for both supernovae and hypernovae in SBGs and star-forming galaxies, 
including the proton diffusion time in the host galaxy and host cluster while undergoing $pp$ interactions. 
These results indicated that SNe and HNe within redshifts $z\siml 4$ could at most provide a fraction
0.2-0.3 of the neutrino background without overproducing the observed gamma-ray background. 
However, there are uncertainties in the star-formation rate at $z\simg 2$, e.g.  \cite{Hopkins+06snrate}, 
as well as in the ratio of HNe to SNe, both of which get worse at higher redshifts. On the other hand, 
as shown by \cite{Chang+16grbnugam}, the $\gamma$-rays from sources at redshifts $\simg 3$ undergo 
increasingly severe degradation due to a rise in the $\gamma\gamma$ interactions in the increasingly 
dense intergalactic photon bath. Thus, the constraints from the Fermi observations can be satisfied 
\cite{Xiao+16nuhn} when one considers a significant contribution of SNe and HNe at redshifts 
$4\siml z \siml 10$ from the first generations of stars (the so-called Population III stars). 
Of course, the apparent surface density of galaxies at high redshifts gets  very large,
which makes it difficult  to correlate any neutrino positions with candidate sources.

A discussion of the general conditions in SBGs and milder star-forming galaxies, including  star formation
rate, gas densities, magnetic fields and dimensions was given by \cite{Lacki+14sfg}. Note  that the  
starburst phenomenon is also suspected, in some cases, to have been initiated by a merger of two galaxies,
in which case large scale shocks would arise which, as discussed in \S \ref{sec:galclu}, leading to CR 
acceleration and secondary neutrinos and $\gamma$-rays. There are also systems in which an AGN and
a starburst co-exist, and based on SBG luminosity functions these could also be relevant for the neutrino
background. Of course, starburst galaxy systems are also subjected to the Fermi-imposed restriction 
requiring effective CR slopes flatter than $\sim -2.2$, e.g. \cite{Murase+13pev,Murase+16uhenurev}.

\subsection{Gamma-Ray Bursts}
\label{sec:grb}

The so-called ``classical" GRBs have long been considered as likely candidates for high energy neutrino 
production \cite{Waxman+97grbnu}. GRBs are catastrophic stellar events brought about by the core collapse 
of a massive star or the merger of a compact degenerate binary, leading to the most energetic explosions 
in the Universe. These result in highly relativistic jets  which emerge from the collapsing or merging
progenitor system, with bulk  Lorentz factors  $\Gamma \sim 10^2-10^3$. In the case of the core-collapse
events (``long" GRBs, with MeV $\gamma$-rays lasting over 2 s) sometimes an accompanying type Ic supernova 
is also detected, in which the progenitor star's envelope is ejected\footnote{However, only a very small 
fraction of all SN Ic  are associated with GRBs.}.
GRBs are detected when spacecraft such as Swift or Fermi trigger on an initial prompt $\gamma$-ray 
burst lasting milliseconds to tens of minutes, over the range of $\sim 0.1-10\MeV$, and sometimes up 
to $\sim 100\GeV$, e.g. \cite{Gehrels+09araa,Meszaros13cta}. The prompt emission is generally followed 
by a slowly decaying afterglow which ranges from X-rays through optical down to radio, over days to months.
The photon spectra of both the prompt and afterglow emission look non-thermal, and have been generally 
ascribed, e.g. \cite{Meszaros13cta}, to electron synchrotron and inverse Compton. The prompt emission 
is typically modeled with Fermi acceleration in internal shocks inside the jet, while the afterglow 
arises from acceleration in an external shock, where the jets plows into the external medium. 
For the prompt emission, besides internal shocks, other alternative mechanisms for the nonthermal emission
have also been proposed, including emission following reconnection or hadronic dissipation at a scattering 
photosphere, or emission from an intermediate zone due to shocks and magnetic reconnection or hadronic 
dissipation and reacceleration, e.g. \cite{Meszaros+00phot,Rees+05photdis,Beloborodov10pn,Zhang+11icmart,
Murase+12reac}. The acceleration of protons is expected to lead, via $p\gamma$ interactions in internal 
shocks, to TeV energy neutrinos \cite{Waxman+97grbnu}, and in external shocks to EeV energy neutrinos 
\cite{Waxman+00nuag}. GeV neutrinos are also expected from proton acceleration and $pp$ or $p\gamma$ 
interactions in photospheres \cite{Murase08grbphotnu,Gao+12photnu,Murase+13subphotnu,Bartos+13pn}, and 
up to multi-TeV can be produced in intermediate magnetic or hadronic dissipation zones 
\cite{Murase+12reac,Asano+14grbcr,Zhang+13grbnu}.

\begin{figure}[h]
\centerline{\includegraphics[width=13cm]{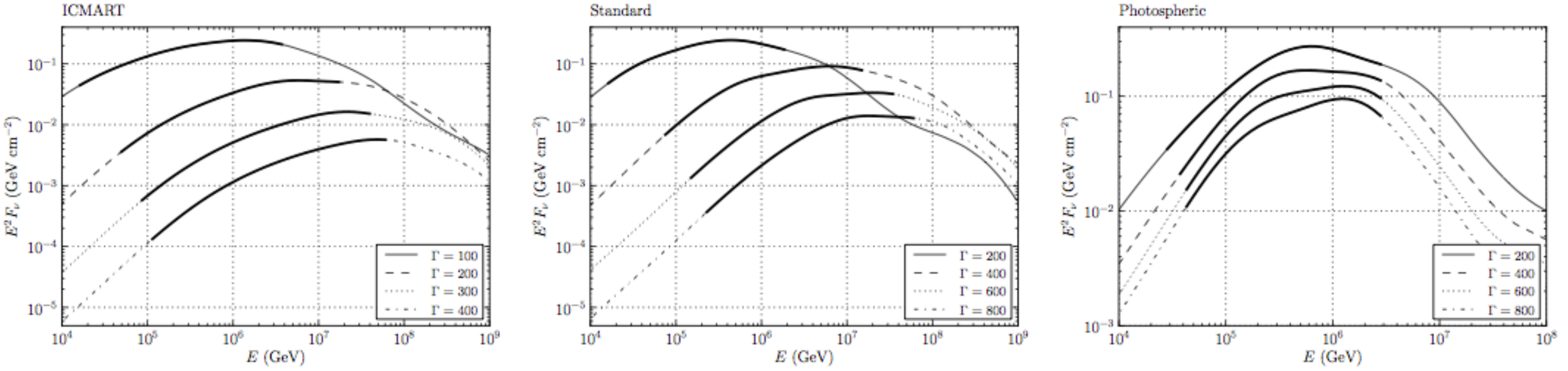}}
\centerline{\includegraphics[width=13cm]{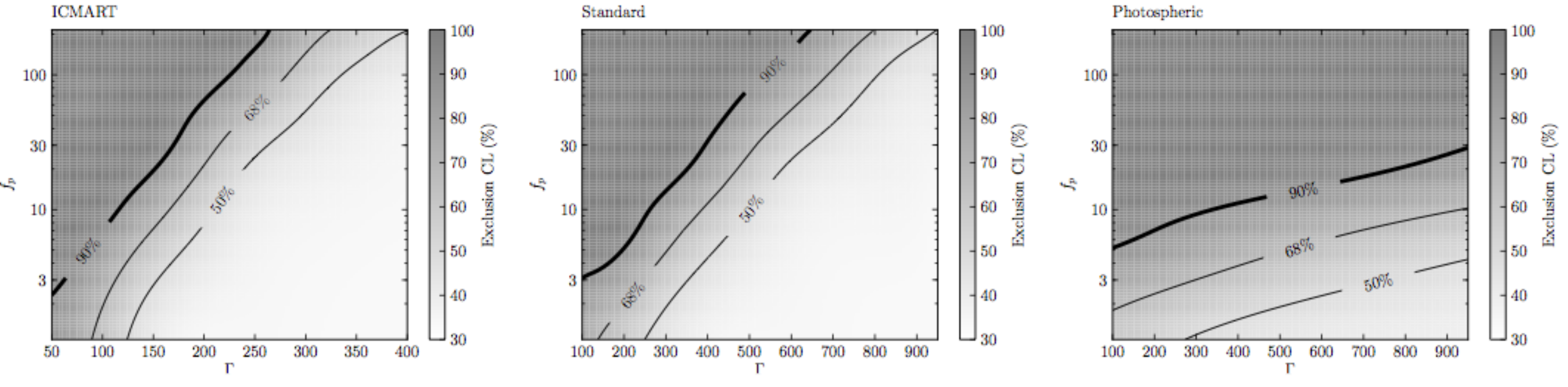}}
\caption{Top: Total normalized neutrino fluxes for ICMART, IS and (baryonic) photosphere models
(left to right) for various Lorentz factors $\Gamma$, scaling with $f_p$ (which here is 10).
Bottom: Allowed region for $f_p$ and $\Gamma$ for the different models. From \cite{IC3+15grbnu4yr}.}
\label{fig:ic3+15grb3lim}
\end{figure}
The initial IceCube tests of GRB neutrino models derived upper limits from the initial 40 string
and 52 string arrays \cite{Ahlers+11-grbcr,Abbasi+11-ic40nugrb,Abbasi+12-IC3grbnu-nat} by comparing
against a simplified internal shock (IS) model with an unchanging radius parameterized by the total 
$\gamma$-ray energy, Lorentz factor $\Gamma$, outflow time variability $t_v$ and a standardized broken 
power law photon spectrum, using the $\Delta$-resonance approximation for the photohadronic interaction, 
and assuming a CR baryon loading (relativistic proton to electron ratio) $f_p=L_p/L_e$.  
This initial study concluded that for $f_p=f_e^{-1}=L_p/L_e=10$ this model over-predicted the data by 
a factor 5, and a model-independent analysis comparing the observed diffuse neutrino flux to that 
expected also gave negative results.  This was an important first result from IceCube, demonstrating 
the ability of a major new Cherenkov neutrino facility to test astrophysical models.
Subsequently, using the same IS model but correcting various approximations and including also 
multi-pion and Kaon channels as well as interactions with the entire target photon spectrum
\cite{He+12nugrb,Li12grbnu,Hummer+12nu-ic3}, lower model fluxes were calculated which did not disagree 
with the 40+56 string data. Interestingly, the original approximate IS calculation of \cite{Waxman+97grbnu} 
had also resulted in a lower flux which is a factor 10 below the WB bound, and within the above IceCube limit.

Much more extensive tests of GRB prompt emission models were made against a set of more accurate internal 
shock models, as well as a magnetic dissipation model (ICMART) \cite{Zhang+11icmart} and a baryonic 
photospheric model \cite{Gao+12photnu,Zhang+13grbnu}, assuming a steady state, fixed radius emission zones.
These statistical tests were made against four years of IceCube data, including two years of the full 
array \cite{IC3+15grbnu4yr}. 
They concluded (Fig. \ref{fig:ic3+15grb3lim}) that at 99\% confidence level less than 1\% of the observed 
diffuse neutrino background can be contributed by the observed sample of 592 EM-detected GRBs.
If the basic acceleration paradigm used for the emission zones is correct, and this result continues to 
stand, it would be indicating that the ratio $f_p=L_p/L_e \siml 1$ in such models. Other photospheric models 
with substantially different neutrino production physics \cite{Murase+13subphotnu,Kashiyama+13pnconv} or 
including time-dependence \cite{Asano+14grbcr} have been calculated which may avoid these restrictions, but 
these have so far been only qualitatively compared against the data.

The classical GRBs discussed above are typically bright, and are EM-detected by spacecraft at an observed 
rate of $\sim 300 \yr^{-1}$, or $\sim 700 \yr^{-1}$ when corrected for viewing constraints, the total sample 
measuring in the thousands.  There are, however, other known or suspected types of GRBs, as discussed next.

\subsection{Low luminosity, shock-breakout, and choked  GRBs}
\label{sec:altgrb}

Low luminosity GRBs (LL GRBs) have been observed for a long time, although only a few are known so far, 
all of which were detected at very close distances $z\ll 1$ due to their intrinsic EM-dimness
\footnote{Unlike classical, high luminosity GRBs, which have been detected in $0.5 \siml z \siml 9$ range.}. 
LLGRBs appear to be a distinct class, although aside from their low luminosity they share many of the 
classical long GRB characteristics, e.g. a non-thermal, albeit softer, spectrum which may be related to a 
relativistic jet which emerged from a collapsing stellar progenitor. Being nearby, a supernova ejecta is 
generally detected as well, which appears to be semi-relativistic, e.g. \cite{Soderberg+06-060218,
Campana+06-060218}. However, their local occurrence per unit volume rate is an order of magnitude higher 
than for classical GRBs, e.g.  \cite{Howell+13-llgrb}.  
classical GRB IS model's neutrino luminosity suggests that LLGRBs could contribute significantly to the 
diffuse neutrino background \cite{Murase+06llgrbnu,Liu+11-llgrbcr}. 

Shock-breakout GRBs, of which even fewer have been detected, also show a soft low luminosity  gamma-ray 
and/or X-ray burst, e.g. \cite{Campana+06-060218}, followed by a brightening of the UV and later optical 
radiation which bears resemblance to a supernova brightening, but with distinct characteristics. 
It is thought that this phenomenon involves a jet emanating from the core of a collapsing massive star,
as for classical long GRBs, but which had less momentum and  just barely managed to break out from the star.
As the jet propagated it imparted  extra energy to the expanding stellar envelope, and the 
boosted supernova shock appears to break out (i.e. the photon diffusion time become shorter than the 
expansion time) in a dense wind which precedes the ejecta \cite{Waxman+07-060218,Chevalier+08break}. 
\begin{figure}[h]
\centerline{\includegraphics[width=13cm]{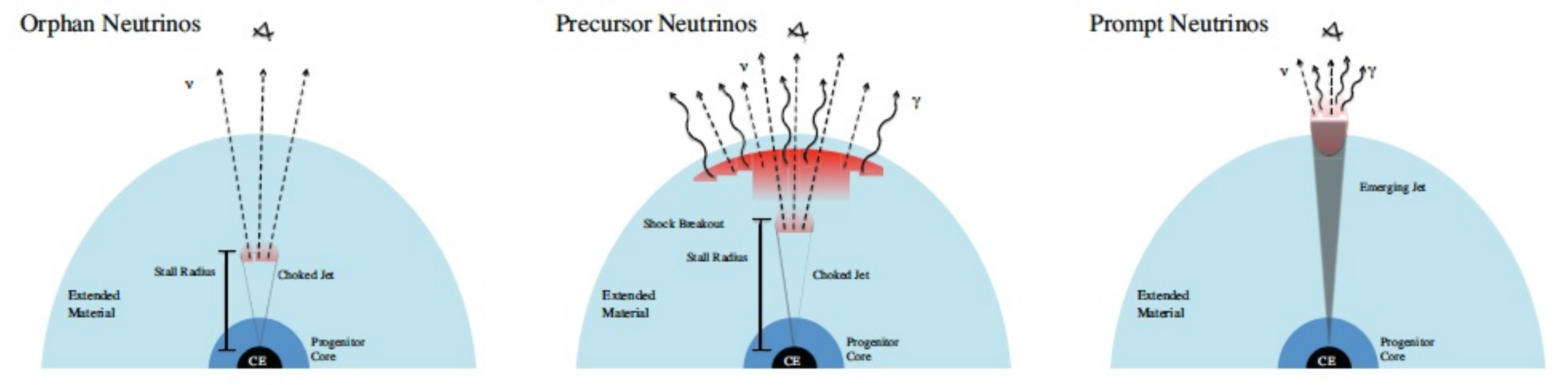}}
\caption{\footnotesize
Sketch of possible scenarios for jet and stellar envelope interaction in a core collapse.
Left: Choked jet and orphan neutrinos. Middle: precursor neutrinos and shock-breakout. Right: Low-luminosity
emergent jet GRB.  From \cite{Senno+16hidden}.  }
\label{fig:Senno+15hidden3jet}
\end{figure}

Choked GRBs, which were posited \cite{Meszaros+01choked}  before shock-breakouts and extragalactic 
neutrinos were discovered, are core-collapse objects where the jets did not emerge, either because they
did not have enough momentum, or because they were not powered long enough to reach the outer radius of 
the stellar envelope.  Internal or recollimation shocks (or magnetic dissipation) in such stalled, buried 
jets could accelerate protons leading to GeV neutrinos, while the gamma-rays would be thermalized, and only 
a subsequent optical supernova would be expected, which at redshifts $z\simg 1$ is rarely detectable. 
Searches with IceCube have so far not yielded candidates \cite{Aartsen+16ic3transchoked}. 
Alternatively, if the jet was energetic enough to eventually emerge, before doing so the pre-emergence 
jet could again undergo shocks or dissipation  giving rise to a neutrino precursor, followed by a successful 
GRB, which could be an LLGRB or a classical GRB as opposed to a failed (choked) GRB \cite{Meszaros+01choked}. 
The shock-breakouts represent an intermediate case between the choked and the emergent. A unified picture 
of the electromagnetic properties expected from all three cases was discussed  by \cite{Nakar15-llgrb}.

All three of these LL GRB types, see Fig. \ref{fig:Senno+15hidden3jet}, would be expected to be ``hidden " 
neutrino sources, since their EM emission is either so weak that only the very few nearest cases trigger 
a $\gamma$-ray detector, or else their EM luminosity is a protracted supernova-like event in the optical/IR, 
and being typically at very high redshift it is again hard to detect. 
Both analytical and in some cases numerical calculations of the high energy neutrino spectral fluences 
were carried out separately for choked jets \cite{Razzaque+03nutomo, Razzaque+04nuhn, Ando+05sngrbnu, 
Horiuchi+08choked, Horiuchi+09choked, Murase+13choked, Fraija15pevchoked}, 
shock-breakout GRBs \cite{Katz+12snbreaknu, Kashiyama+13breakout, Giacinti+15nusn}  and LLGRBs 
\cite{Murase+06llgrbnu,Gupta+07grbnu}. 

\begin{figure}[h]
\begin{minipage}{0.5\textwidth}
\centerline{\includegraphics[width=2.8in]{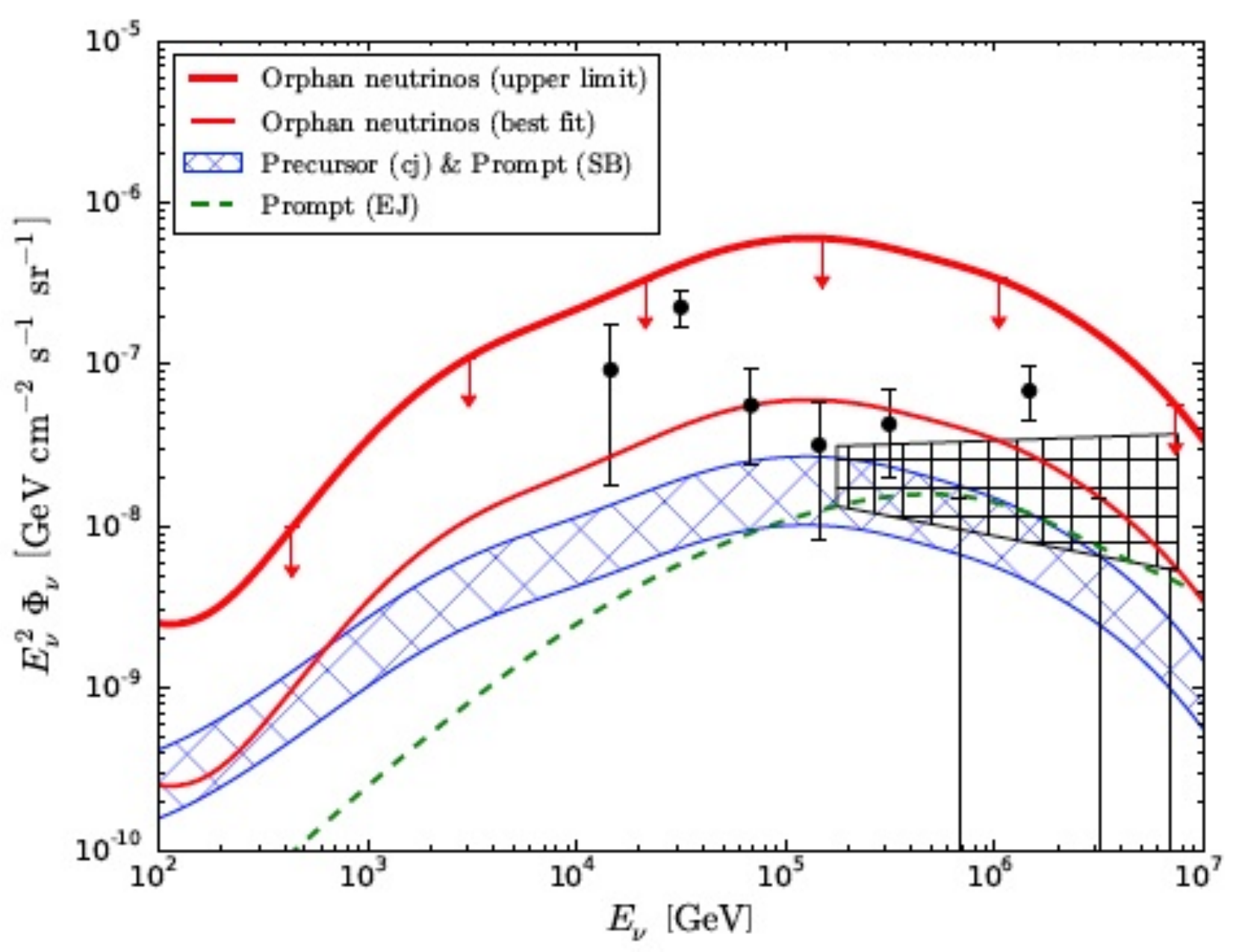}}
\end{minipage}
\hspace{5mm}
\begin{minipage}[t]{0.50\textwidth}
\vspace*{-0.9in}
\caption{\footnotesize
Predicted all-flavor diffuse neutrino fluxes from three types of low-luminosity GRBs: choked jets (orphan 
neutrinos, in red); precursor and shock-breakout neutrinos (blue); and prompt emergent jet LLGRB neutrinos 
(dashed).  From \cite{Senno+16hidden}.  }
\label{fig:Senno+15hiddenfg4}
\end{minipage}
\end{figure}

More recently, a unified calculation and comparison of all three types of hidden GRBs (choked, break-out 
and low-luminosity) was carried out \cite{Senno+16hidden}, see Fig. \ref{fig:Senno+15hiddenfg4}. 
This calculation used a standard GRB luminosity function, and neutrino emission was considered only from choked 
or precursor jets whose luminosity was low enough to ensure that the buried shocks are not radiation-broadened
(since buried jets are at lower radii and their radiation density is comparatively higher than in emergent jets).
This ensures a collisionless shock, in which that particles scattered between pre- and post-shock regions 
are subjected to the full bulk velocity difference, as needed for first-order Fermi acceleration. 
Otherwise, for higher luminosity buried jets, the photon mean-free path governs the shock width, which becomes 
larger than the typical photon mean-free path or gyro-radius, the scattered particles are not exposed to 
the full bulk velocity difference, and classical first-order Fermi acceleration is not expected, e.g.
\cite{Levinson+08radshock,Murase+13choked}. Low-power jets are also required in order for the jet
to stall before it emerges from the star \cite{Murase+13choked,Senno+16hidden}.

\noind
The conclusion from this calculation \cite{Senno+16hidden} is that a combination of choked jet, shock-breakout 
and low-luminosity GRBs could in principle provide the observed IceCube neutrino flux, without violating 
either the Fermi observations nor the (classical GRB) stacked neutrino analyses.

\subsection{Other sources: tidal disruptions, white dwarf mergers}
\label{sec:other}

Tidal disruption (TDE) events of stars by massive black holes at the centers of some galaxies can
also lead, in a fraction of the cases, to relativistic jets, as in the gamma-ray source Sw J1644+57 
\cite{Burrows+11-1644tid}. It has also been proposed that TDEs could be accelerators of UHECRs
\cite{Farrar+14tiduhecr}. If such tidal disruptions occur in a galactic whose bulge gas density is 
large enough, or if the tidal disruption initially leads to a precursor wind before the jet is produced, 
e.g. \cite{Metzger+16tid}, this external gas may either choke the jet or it may lead to a shock-breakout 
similar to that in GRBs, e.g. \cite{Wang+16tdecrnu}. 
The rates are highly uncertain, but a fraction of the observed VHE neutrinos may arise from such events,
whose gamma-rays could be effectively EM-hidden because of the high optical depth of the enshrouding gas,
some recent calculations being, e.g., \cite{Senno+16tde,Dai+17tde,Lunardini+17tde}.

White dwarf (WD) mergers are another possible type of hidden neutrino source. WDs are the remnants of 
most SN explosions, and WD binaries are abundant enough that their merger rate is estimated to be
comparable to that of the SN Ia. Such WD mergers may lead to a magnetized outflow \cite{Beloborodov14wdmerg}, 
in which photons are trapped up to the diffusion radius where the diffusion time is equal to the dynamic 
time.  Magnetic reconnection in the flow beyond the diffusion radius can lead to proton acceleration which 
gives rise to $pp$ interactions resulting in secondary neutrinos and gamma-rays \cite{Xiao+16wdmnu}.
Since the scattering optical depth is still large at the diffusion radius, the gamma-rays are 
degraded and these sources are effectively dark, or at least considerably dimmed, as far as the Fermi
energy sensitivity range, thus avoiding the Fermi constraint. The neutrino flux can be a substantial 
fraction of the IceCube flux, depending on model uncertainties and WD merger rates. 

There are other stellar sources in starburst galaxies which have been considered for producing 
very high energy neutrinos, including magnetars, young pulsars and macro-novae, which cannot be
covered here.

\section{Discussion}
\label{sec:disc}

The discovery of extragalactic very high energy neutrinos by IceCube has opened an entirely new
realm of possibilities for exploring the physics of the highest energy astrophysical sources, 
potentially out to the most distant reaches of the Universe. However, the small number of events 
at these energies allows one to address so far only the aggregate emission in the form of 
a diffuse background radiation. The identity and nature of the sources remains unknown, although
it is realistic to expect significant progress in this respect with multi-messenger approaches, such 
as the AMON project \cite{Smith+13amon,Cowen+16amon} and others, especially those involving neutrino 
detections combined with the (relatively) more easily detectable electromagnetic counterparts.

With accumulating high energy neutrino and gamma-ray data it will become increasingly feasible to 
draw general conclusions about the physical mechanisms producing the neutrinos, as well as about 
the general environment in which they originate and in which their secondaries propagate, e.g. 
\cite{Aartsen+15-IC3nubkggmaxlik}. Furthermore, these neutrinos and their co-produced gamma-rays must 
be linked to high energy cosmic rays of energy in the $10^{14}-10^{17}\eV$ range, and possibly beyond.
The fact that the IceCube neutrino emission level is close to the Waxman-Bahcall limit 
\cite{Waxman+99bound} has provide dthe  motivation for an interesting argument \cite{Katz+13uhecr} 
indicating that the input of cosmic ray energy per decade over their entire  spectral range may be 
approximately constant at the level of $\sim 10^{44}\erg~\Mpc^{-3}\yr^{-1}$, whose manifestation 
in the IceCube range would be the observed neutrinos, e.g. \cite{Waxman15sbgnu,Murase+16uhenurev}.

Further progress can be expected with the future completion of the KM3NeT under-water neutrino 
detector in the Mediterranean \cite{KM3NeT+16loi}, with roughly similar capabilities as IceCube
and a complementary northern hemisphere location. Both IceCube \cite{IC3+15Gen2} and KM3NeT 
\cite{KM3NeT+16loi} have proposed extension proposals to their sensitivity to lower and higher energies, 
which could also address interesting questions of fundamental physics, e.g. \cite{Shoemaker+16ic3bsm},
and dark matter \cite{IC3+16solardm3,IC3+16galaxydm}. Much larger effective area detectors, such as 
the ANITA balloon telescope, e.g. \cite{Vieregg+12anita2nu}, the proposed very large radio arrays
in Antarctica such as ARIANNA \cite{Barwick+16arianna} and ARA \cite{Besson+15aranulim}, and
space-based detectors such as JEM-EUSO \cite{JEM-EUSO+16}, will extend the sensitivity to the very
low fluxes expected at energies in the $10^{20}-10^{21}$ eV range and above, which can address 
important questions of cosmogenic neutrino production, including testing for the presence of UHECR at 
or beyond the GZK radius $\sim 100\Mpc$, whether the spectrum extends beyond the GZK energy 
$\sim 6\times 10^{19}\eV$, constraining the heavy element content, etc.
The completion of the approved CTA (Cherenkov Telescope Array) large ground-based VHE gamma-ray 
detector, e.g.  \cite{Bigiongiari16-CTA}, will also be extremely useful for simultaneous neutrino
and gamma-ray detections, localizations and source characterizations.
\\

%
\noind
{\bf MAIN SUMMARY POINTS:}
\begin{enumerate}
\item Extragalactic TeV-PeV neutrinos have been discovered, heralding a completely new channel for 
  studying extreme high energy cosmic physical processes at the highest redshifts in the  Universe.
\item These neutrinos carry important clues for investigating the origin of the high energy cosmic 
  rays, and they provide stringent constraints for the possible source models being considered.
\item A smoking-gun identification of the actual sources remains so far elusive, since the angular accuracy 
  of the arrival directions of individual neutrinos remains of the order of a degree or larger.
\item Co-emitted gamma-rays, when detected, are likely to help address this problem, as will
  also statistical analyses based on larger numbers of neutrinos and more complete candidate source 
  catalogs.
\item We live in an exciting new era, where tremendous progress is being made.
\end{enumerate}

\noind
{\bf FUTURE ISSUES}
\begin{enumerate}
\item The main desiderata for future advances will be to achieve significantly higher event statistics, 
  which requires, e.g., the approval and building of the Gen2 extensions to IceCube, the completion of new 
  facilities such as KM3NeT, and the expansion of multi-messenger localization operations such as AMON.
\item The above sensitivity increases are also crucial for investigating pressing issues of basic neutrino 
  physics, for dark matter searches, and other beyond-the-standard model questions.
\item A major current problem is  that the observed diffuse neutrino background flux appears to 
 over-predict the observed diffuse gamma-ray flux, assuming the `usual suspect' optically-thin candidate sources. 
\item Unless otherwise resolved, the above issue is suggestive of  electromagnetically hidden sources, 
  i.e. sources where gamma-rays are absorbed or degraded. Possibilities being considered are
  buried low-luminosity GRB jets or tidal disruption jets, among others, but much more work remains 
  to be done.
\item Many new surprises can be expected from the major new facilities coming online in the next decade.
\end{enumerate}

\section*{DISCLOSURE STATEMENT}
The author is not aware of any affiliations, memberships, funding, or financial holdings that
might be perceived as affecting the objectivity of this review. 

\section*{ACKNOWLEDGMENTS}
The author is grateful to Douglas Cowen, Kohta Murase and Marek Kowalski for useful communications.
and to NASA NNX13AH50G for partial support.



\section{Appendix}
\label{sec:app}

\subsection{Neutrino Production Mechanisms}
\label{sec:nuprod}

Aside from the possibility of high energy ($\simg$ GeV) neutrinos being produced by the decay of
exotic (beyond the Standard Model) particles, the production of such neutrinos is expected in
astrophysical scenarios. 
Cosmic rays can lead, via hadronic interactions such as $pp$, $pn$ or $p\gamma$ to the production 
of mesons, mainly  $\pi^\pm, \pi^0$, which decay as, e.g., $\pi^+ \to  \mu^+\nu_\mu$  followed by 
$\mu^+ \to e^+ \nu_e {\bar \nu}_\mu$ or $\pi^0 \to 2\gamma$. For proton (or neutron) CRs the 
energy of the decay neutrinos is typically related to the parent cosmic ray $p$ or $n$ energy
by $\vareps_\nu \simeq 0.04-0.05 \vareps_{p,n}$. 
For a total cosmic ray (proton) volumetric energy generation rate $Q_{p}~[\erg~\Gpc^{-3}\yr^{-1}]$ 
leading to a CR energy generation rate per decade of energy $\vareps_p Q_{\vareps_p}$, one expects 
an all-flavor neutrino energy generation rate per decade of energy of
\beq
\vareps_\nu Q_{\vareps_\nu} \approx 
    \frac{K}{(1+K)} \left(\frac{3}{4}\right) \min [1,f_{p\gamma/pp}] \vareps_p Q_{\vareps_p},
\enq
where by $\vareps_j$ we denote the source-frame energy of particles of type $j$.
Here the (3/4) factor enters because roughly 1/4 of the energy in the decay chain is lost to $e^\pm$ 
which end going into photons, and $K$ is the average number ratio of charged to neutral pions, which is
$K\simeq 1$ for the $p\gamma$ and $K\simeq 2$ for $pp,pn$ processes. The factor $\min [1,f_{p\gamma/pp}]$ 
is the $pp$ or $p\gamma$ meson production efficiency,
\beq
f_{p\gamma/pp} \simeq n_{\gamma /p} \kappa_p \sigma_{p\gamma /pp}^{incl} c t_{int}
\enq
where $n_{\gamma /p}$ is the number density ot target photons (protons), $\kappa_p$ is the
inelasticity (i.e. the relative energy loss per interaction, $\Delta \vareps_p/\vareps_p$),
which on average is $\kappa_p \simeq 0.5$  for both $pp$ and $p\gamma$, the inclusive 
(i.e. total) cross section is $\sigma_{p\gamma}^{incl} \simeq 5\times 10^{-28}\cm^2$ for $p\gamma$, 
or $\sigma_{pp}^{incl}\simeq 8\times 10^{-26}\cm^2$ for $pp$, and $t_{int}$ is the
time available for interactions, with $ct_{int}$ the interaction length. The interaction 
time is generally $t_{int}=\min[t_{inj}, t_{esc}, t_H]$, where $t_{inj}$ is the CR injection 
time, $t_{esc}$ is the CR escape time from the interaction region, $t_H$ is the local Hubble 
time or age of the Universe in the source frame, and the interaction length is along a 
random walk path between interactions.
 
The neutral pions result in an accompanying gamma-ray emission, $\pi^0 \to 2\gamma$,
which is related (at the source) to the neutrino emission by
\beq
\vareps_\gamma Q_{\vareps_\gamma} 
  \approx \frac{1}{K}\frac{4}{3} \left( \vareps_p Q_{\vareps_p}\right) |_{\vareps_\nu=\vareps_\gamma /2}
\enq
where the energy of the $\gamma$-rays are, on average, $\vareps_\gamma \simeq 2\vareps_\nu$.

Denoting with $E_j$ the energies of particles $j$ observed at Earth, the neutrinos observed 
by IceCube from a source at redshift $z$ are related to the parent proton energy by
\beq
E_\nu \sim 0.05 E_p \simeq 2~{\rm PeV}~\vareps_{p.17} \left[2/(1+z)\right],
\enq
and the all-flavor diffuse neutrino flux $\Phi_\nu$ per steradian with observed energy $E_\nu$ at Earth is
\beq
E_\nu^2 \Phi_\nu = 
\frac{c}{4\pi}\int\frac{dz}{(1+z)^2 H(z)}\left[\vareps_\nu Q_\nu\right]|_{\vareps_\nu=(1+z)E_\nu},
\label{eq:inbex}
\enq
in units of, e.g. $\GeV\cm^{-2}\s^{-1}\sr^{-1}$. 
Here $H(z) \simeq H_0 \left[\Omega_V + (1+z)^3\Omega_M\right]^{1/2}$ is the redshift-dependent 
Hubble parameter, with $H_0\simeq 70$ Km/s/Mpc.

For a local ($z=0$) CR proton differential energy input rate $Q_{E_p}$ the  diffuse neutrino
background {\it per flavor} at Earth which follows from eq.(\ref{eq:inbex}) is approximately
\beq
E_\nu^2 \Phi_\nu \approx \frac{c t_H \xi_z}{4\pi} \left[\frac{K}{4(1+K)}\right]
    \min[1,f_{p\gamma /pp}] \left(E_p Q_{E_p}\right),
\label{eq:inbapprox}
\enq
e.g. \cite{Waxman+97grbnu,Murase+13pev}, where the per-flavor factor is $[K/4(1+K)]=[1/8,~1/6]$.
i.e. 1/8 for $p\gamma$ ($K=1$) or 1/6 for $pp$ ($K=2$), $t_H\simeq 13.2$ Gyr, and $\xi_z$ is a 
redshift evolution factor, which, e.g. for sources evolving approximately as the star-formation 
rate, such as GRBs or SNe, is $\xi_z\sim 3$ at $z\sim 1$.
The corresponding diffuse $\gamma$-ray background associated with eq.(\ref{eq:inbapprox}),
in the absence of electromagnetic (EM) cascades, is given by
\beq
E_\gamma^2 \Phi_\gamma \approx 2 \left( E_\nu^2 \Phi_\nu \right)|_{E_\nu=0.5 E_\gamma}
\label{eq:igbapprox}
\enq

\subsection{Cosmic Ray Acceleration Mechanisms}
\label{sec:acc}

The photons or hadrons entering the $p\gamma$ or $pp$, $pn$, etc.  interactions 
must have energies such that the center of momentum (CM) energy is above threshold for producing 
pions or other mesons. Thus either the photons involved in $p\gamma$ must have high lab-frame energies, 
or the the hadrons initiating the $pp$ or $p\gamma$ interactions must be highly relativistic, i.e. they
must be cosmic rays (CRs). Among the most promising mechanisms for CR acceleration are the diffusive 
shock acceleration (DSA) and magnetic reconnection acceleration, both of which are first-order Fermi 
type mechanisms; stochastic or turbulent acceleration, which is a second-order Fermi type; and 
electrostatic type acceleration mechanisms, such as pulsar magnetospheric acceleration, or wake field
acceleration.

The DSA mechanism can arise in systems where a strong shock propagates, with charged particles 
scattering back and forth across the shock interface. Typically such shocks are collisionless, 
i.e. the binary particle collision mean-free-path is very large compared to that for scattering 
by magnetic irregularities. For a sub-relativistic (or mildly relativistic) shock propagating 
into a stationary upstream medium, a proton which is already relativistic in the downstream (moving) 
region can run ahead of the shock, and will randomized by scattering againts magnetic irregularities 
in the upstream  region The particle is then overtaken 
by the shock, finding itself again in the downstream region where it is again randomized again by 
magnetic irregularities. The process then repeats itself, and at each step of the cycle the 
particle gains energy at the expense of the upstream-downstream gas relative bulk velocity 
difference. Each time the particle is hit head-on, the  net relative energy boost being
$\Delta E/E \propto (v_s/c)$ (hence first order), where $v_s$ is the shock velocity\footnote{Two
things to note: one is the initial particles injected must already be at least mildly relativistic; 
and for relativistic shocks, the treatment is more complicated after the first scattering, but
under some approximations qualitatively similar results are expected.}.

In such first order shock acceleration scenarios, the typical acceleration timescale is
\beq
t_{ac}^{dsa}\simeq \eta  \left(\frac{r_L}{c}\right) \beta_s^{-2}
\label{eq:tacdsa}
\enq
where $r_L=E/ZeB$ is the Larmor radius, $Ze$ is particle charge, $B$ is magnetic field and
$\beta_s=v_s/c$ is the shock velocity in the upstream frame, and $\eta \sim 1-10$. This acceleration
timescale is of the order of the gyration time, being proportional to the maximum particle energy 
$\vareps$ reached, and is controlled by the {\it spatial} diffusion time.
If the shocks occur in a jet oriented towards the observer with a bulk Lorentz factor $\Gamma$, the 
acceleration time, Larmor radius, magnetic field and shock velocity in eq.(\ref{eq:tacdsa}) should 
be read as the corresponding quantities in the jet frame, $t'_{ac}$, $r'_L$, $B'$ and $\beta'_s$. 
A limit on the maximum energy is imposed by requiring that $t_{ac}$ not exceed 
the dynamic time, $t_{dyn}\simeq R/v_s$ in the non-relativistic (NR) case, where $R$ is
the lab-frame dimension of the acceleration region (e.g. radius of shock),
\beq
E_{max}\simeq \left(\frac{\beta_s}{\eta}\right) \frac{Ze B R }{(1+z)},
\label{eq:emaxdsa}
\enq
where $z$ is the redshift of the source\footnote{More exactly, if the minimum spatial diffusion 
coefficient in the Bohm limit is $D_{min}=\eta' r_L c/3$, then $t_{ac}^{dsa}\simeq (20\eta'/3)
(r_L/c)\beta_s^{-2}$, and $\vareps_{max}\simeq (3\beta_s/20\eta') ZeBR/(1+z)$}. 
This criterion is equivalent to the confinement criterion that $r_L$ be smaller than the acceleration 
region $R$. 
For shocks in a jet with bulk Lorentz factor $\Gamma$, in the comoving frame $t'_{dyn}\simeq R/c \Gamma$, `
so $(20\eta/3) (r'_L/c) {\beta'_s}^{-2} \leq (R/c\Gamma)$ leads to a lab-frame maximum energy given by
eq.(\ref{eq:emaxdsa}) but with $\beta_s$ replaced by $\beta'_s \to 1$ and $B$ replaced by $B'$, the 
comoving field. Alternatively, the maximum energy may be limited by the requirement that the acceleration 
time be shorter than the synchrotron radiation loss time of the accelerated particle.

Magnetic reconnection is another acceleration process which operates as a Fermi first order mechanism.
Long considered as the cause of particle acceleration in solar flares, its occurrence is
expected to be ubiquitous in many astrophysical situations where shear, turbulence or rotation lead 
to reconnection. Candidate sites include, besides flares, azimuthally
sheared accretion disks, transverse shear between jets and environment, MHD turbulent media, etc.
A` schematic X-point geometry considers regions of dimension $\ell_{rec}$ of opposite magnetic polarity 
which approach each other, e.g. along the $\pm y$ direction, at a sub-relativistic speed $\beta_{rec} 
\siml 0.1$ leading to a thin reconnection layer with an electric field along the $x$-axis where 
plasma flows out along the $\pm x$-axis at the Alfv\'en speed $V_A\sim 1$. Charged particles are
caused to rotate repeatedly in and out of the opposite converging regions under the action of the
opposite magnetic field polarities, while experiencing a net acceleration along the $x$-axis  under
the effect of the reconnection layer's electric field. A simple but illustrative calculation 
\cite{Giannios10crmag} shows that the effective acceleration timescale is
$t_{ac}^{rec}\simeq ({2\pi}/{[1-1/A]}) ({r_L}/{c}) \sim 4\pi ({r_L}/{c})$,
where for reasonable reconnection rates $A\sim 2$, giving an acceleration timescale which is essentially 
eq.(\ref{eq:tacdsa}) for the diffusive shock acceleration, i.e. roughly the gyration period.  The maximum 
particle energy is again obtained by equating the acceleration time to the dynamic time, leading 
approximately to eq.(\ref{eq:emaxdsa}), or by equating the acceleration time to the radiation loss time. 

Stochastic acceleration, such as expected in MHD turbulent media as particles are scattered by
waves of velocity $v_w$ with random  orientations, leads as mentioned to relative energy changes 
which are second order, $\propto (v_w/c)^2$, because the particles suffer both head-on and overtaking 
collisions with the waves. This is a process of diffusion in energy space, the particles sometimes 
gaining ans sometimes losing energy, the first order energy changes canceling out, but resulting on
a net average energy gain.  For magnetic field fluctuations with a spectral energy density
$W_k \propto k^{-q}$, where $k=$ wavenumber corresponding to turbulent lengthscales $\ell \sim 2\pi/k$,
for resonant scattering (where $r_L\sim k$) one expects an energy diffusion coefficient
$D_{\vareps\vareps}\propto \vareps^q$, and a scattering and acceleration time $t_{ac}\propto \vareps^{2-q}$,
\beq
t_{ac}^{sto}\simeq \frac{\vareps^2}{4 D_{\vareps\vareps}}
            \sim \eta_{sto} \frac{\ell_{t}}{c}\left(\frac{r_L}{\ell_{t}}\right)^{2-q}
\label{eq:taccsto}
\enq
where $\ell_{t}\sim 2\pi/k_{min}$ and $\eta_{sto}\sim 1$, e.g. \cite{Dermer+96stochacc,Bykov+96slowheat,
Petrosian+04stochacc}.  This timescale is generally longer than that of eq.(\ref{eq:tacdsa}) for shock 
acceleration, but is still shorter than the MHD wave timescale $\ell_{t}/c$, so it can act as a 
slow-heating mechanism on a sub-hydrodynamical timescale \cite{Bykov+96slowheat,Murase+12reac}.  
At the highest energies, where $r_L\sim \ell_{t}$, or for values of $q=2$ as suggested by various MHD 
turbulence simulations, the energy diffusion coefficient becomes $D_{\vareps\vareps}\propto \vareps^2$ 
\cite{Brunetti+07stochacc}, and the stochastic timescale becomes comparable to eq.(\ref{eq:tacdsa}) 
for shock acceleration. Equating the acceleration time (\ref{eq:taccsto}) to the hydrodynamic time
$R/c\beta\Gamma$  (or $R/u=R/c\beta$ in the non-relativistic case) the lab-frame maximum energy is
\beq
E_{max}\simeq Ze B\ell_t \times
            \cases{
            1                                     &~~~{\rm for}~~ $q=2$ ; \cr
            (R/\eta \beta\Gamma \ell_t)^{1/(2-q)} &~~~{\rm for}~~ $q\neq 2$
}
\label{eq:emaxsto}
\enq
where for a jet the comoving $B'$ value should be used. 
\\
~
\\
\hspace*{7cm}



\footnotesize

\bibliographystyle{ar-style5.bst}



\end{document}